\documentclass[english]{article}
\usepackage[T1]{fontenc}
\usepackage[latin9]{inputenc}
\usepackage[active]{srcltx}
\usepackage{amsmath}
\usepackage{amssymb}
\usepackage{stackrel}
\usepackage{graphicx}

\makeatletter
\newcommand{\lyxaddress}[1]{
\par {\raggedright #1
\vspace{1.4em}
\noindent\par}
}

\makeatother

\usepackage{babel}
\begin{document}

\title{\textbf{Quartic time-dependent oscillatons}}

\date{A. Mahmoodzadeh\thanks{a.mahmoodzadeh@iau-boukan.ac.ir}, B. Malekolkalami\thanks{B.Malakolkalami@uok.ac.ir }}
\maketitle

\lyxaddress{\textit{Faculty of Science, University of Kurdistan, Sanandaj, P.O.Box
416, Iran}}
\begin{abstract}
In this paper, we will study some properties of oscillaton, spherically
symmetric object made of a real time-dependent scalar field, Using
a self-interaction quartic scalar potential instead of a quadratic
or exponential ones discussed in previous works. Since the oscillatons
can be regarded as models for astrophysical objects which play the
role of dark matter, therefore investigation of their properties has
more importance place in present time of physics$^{,}$ research.
Therefore we investigate the properties of these objects by Solving
the system of differential equations obtained from the Einstein-Klein-Gordon
(EKG) equations and will show their importance as new candidates for
the role of dark matter in the galactic scales.\bigskip{}
\end{abstract}
\textbf{I. INTRODUCTION}

\bigskip{}

The first evidence for dark matter appeared in the 1930s, when astronomer
Fritz Zwicky noticed that the motion of galaxies bound together by
gravity was not consistent with the laws of gravity. Zwicky argued
that it should be more matter than what is visible and he named \textit{dark
matter} this unkown kind of invisible matter. Since that time, numerous
evidence has confirmed the existence of dark matter. For instance,
Galaxy rotation curves, galaxy cluster composition, bulk motions in
the Universe, gravitational lensing, the formation of large scale
structure (LSS) and red shift are examples that prove there should
be more than just visible matter in the Universe, what is in the form
of invisible matter or non-baryonic. Unfortunately one of the biggest
challenge in astrophysics and particle physics which has remained
unsolved as of yet, is the problem of dark matter. Therefore the number
of proposals have been presented for solving the problem of dark matter
theory have had an increasingly process in recent years. Standard
model particle physics, including WIMPs, super WIMPs, light gravitinos,
hidden dark matter, sterile neutrinos, axions, and other models based
on warm dark matter, particles with self interactions, complex scalar
field for bosonic dark matter were therefore under scrutiny, Although
efforts have not yielded a unit and certain result so far {[}1-4{]}.
Nowadays the another alternative which has paid much attention, is
real scalar field and study of oscillatons made of a real time-dependent
scalar field for solving the hypothesis of dark matter has found general
importance at galactic scales {[}5-6{]}.

\vspace{1cm}

\textbf{II. MATHEMATICAL BACKGROUND\bigskip{}
}

In this section, we study the case of a self-interaction qurtic scalar
potential with spherical symmetry similar to what analyzed in {[}7{]}.
The most general spherically-symmetric metric case is written as

\begin{flushleft}
\hspace{30bp}$ds^{2}=g_{\alpha\beta}dx^{\alpha}dx^{\beta}=-e^{\nu-\mu}dt^{2}+e^{\nu+\mu}dr^{2}+r^{2}(d\theta^{2}+sin^{2}\theta d\varphi^{2})$,\hspace{25bp}(1)
\par\end{flushleft}

where $\nu=\nu(t,r)$ and $\mu=\mu(t,r)$ are functions of time and
spherical radial cordinate (we have used natural units in which \textit{$c=1$}).
Tensor for a real scalar field $\Phi(t,r)$ with a scalar potential
field $V(\Phi$) is defined as {[}7, 8, 9{]}.

\begin{flushleft}
\hspace{80bp}$T_{\alpha\beta}=\Phi_{,\alpha}\Phi_{,\beta}-\frac{1}{2}g_{\alpha\beta}[\Phi^{,\gamma}\Phi_{,\gamma}+2V(\Phi)]$.\hspace{83bp}(2)
\par\end{flushleft}

\begin{flushleft}
The non-vanishing components of $T_{\alpha\beta}$ are
\par\end{flushleft}

\begin{flushleft}
\hspace{68bp}$-T^{0}$$_{0}=\rho_{\Phi}=\frac{1}{2}[e^{-(\nu-\mu)}\dot{\varPhi^{2}}+$$e$$^{-(\nu+\mu)}\Phi^{\prime2}+2V(\Phi)]$,\hspace{47bp}(3)
\par\end{flushleft}

\begin{flushleft}
\hspace{70bp}$T_{01}=p_{\Phi}=\dot{\varPhi}\varPhi^{\prime}$,\hspace{190bp}(4)
\par\end{flushleft}

\begin{flushleft}
\hspace{70bp}$T_{1}^{1}=p_{r}=\frac{1}{2}[e^{-(\nu-\mu)}\dot{\varPhi^{2}}+e^{-(\nu+\mu)}\Phi^{\prime2}-2V(\Phi)]$,\hspace{55bp}(5)
\par\end{flushleft}

\begin{flushleft}
\hspace{70bp}$T_{2}^{2}=p_{\bot}=\frac{1}{2}[e^{-(\nu-\mu)}\dot{\varPhi^{2}}-e^{-(\nu+\mu)}\Phi^{\prime2}-2V(\Phi)]$,\hspace{53bp}(6)
\par\end{flushleft}

and we have also $T^{3}$$_{3}=T^{2}$$_{2}$. Overdots denote $\frac{\partial}{\partial t}$
and primes denote $\frac{\partial}{\partial r}$. The different components
mentioned above are identified as the energy density, $\rho_{\varPhi}$,
the momentum density, $p_{\varPhi}$, the radial pressure, $p_{r}$
, and the angular pressure, $p_{\bot}$, respectively. Einstein equations,
$G_{\alpha\beta}=R_{\alpha\beta}-\frac{1}{2}g_{\alpha\beta}R=k_{0}T_{\alpha\beta}$
are used to obtain differential equations for functions $\nu,\mu$
then

\begin{flushleft}
\hspace{110bp}$(\nu+\mu)^{.}=k_{0}r\dot{\varPhi}\Phi^{\prime},$\hspace{140bp}(7)
\par\end{flushleft}

\begin{flushleft}
\hspace{110bp}$\nu^{\prime}=\frac{k_{0}}{2}(e^{2\mu}\dot{\varPhi^{2}}+\Phi^{\prime2})$,\hspace{127bp}(8)
\par\end{flushleft}

\begin{flushleft}
\hspace{108bp}$\mu^{\prime}=\frac{1}{r}[1+e^{\nu+\mu}(k_{0}r^{2}V(\Phi)-1)]$,\hspace{82bp}(9)
\par\end{flushleft}

where $R_{\alpha\beta}$ , $R$ are the Ricci tensor and Ricci scalar
respectively and $k_{0}=8\pi G=\frac{8\pi}{m_{pl}^{2}}$ . The universal
gravitational constant, $G$ , is the inverse of the reduced Planck
mass squared $m_{pl}$. The conservation equations for the scalar
field energy- momentum tensor (2) requires to have

\begin{flushleft}
\hspace{102bp}$T_{;\beta}^{\alpha\beta}=[\square\Phi-\frac{dV(\Phi)}{d\Phi}]\varPhi^{,\alpha}=0$,\hspace{100bp}(10)
\par\end{flushleft}

where $\square=\partial_{\alpha}\partial^{\alpha}=g_{\alpha\beta}\partial^{\alpha}\partial^{\beta}$
is the d$^{,}$Alembertian operator. Therefore we can obtain the Klein-Gordon
(KG) equation for the scalar field $\Phi(t,r)$

\begin{flushleft}
\hspace{70bp}$\Phi^{\prime\prime}+\Phi^{\prime}(\frac{2}{r}-\mu^{\prime})-e^{\nu+\mu}\frac{dV(\Phi)}{d\Phi}=e^{2\mu}(\overset{..}{\Phi}+\overset{.}{\mu}\overset{.}{\Phi})$.\hspace{60bp}(11)
\par\end{flushleft}

As we can see this differential equation, is fully related to scalar
potential field and is considered as the representative of all cases
of oscillatons with any kind of $\Phi(t,r)$ and$V(\Phi)$ {[}8{]}
.\vspace{1cm}

\textbf{III. QUARTIC POTENTIALS\bigskip{}
}

The hypothesis of scalar dark matter in the universe with a minimally
coupled scalar field and a scalar potential in the form of quadratic,
exponential or $cosh$ has been discussed before {[}7, 9, 10{]}. But
in this study we are interested in to investigate the self- interaction
of an oscillaton only, which is described by a quartic form of scalar
field for the role of dark matter at the cosmological scale. This
scalar field potential can be written as

\hspace{132bp}$V(\Phi)=\frac{1}{4}\lambda\varPhi^{4}$,\hspace{116bp}(12)

where $\lambda$ is the quartic interaction parameter which is obtained
through constraints imposed on formulation of the problem. If we choose
$\Phi(t,r)$=$\sigma(r)\phi(t)$ , then equation (11) reads as

\begin{flushleft}
\hspace{48bp}$\phi\{\sigma^{\prime\prime}+\sigma^{\prime}(\frac{2}{r}-\mu^{\prime})\}-\lambda e^{\nu+\mu}\varPhi^{3}=e^{2\mu}\sigma(\ddot{\phi}+\overset{.}{\mu}\dot{\phi})$.\hspace{72bp}(13)
\par\end{flushleft}

Taking into account with the Fourier expansion

\begin{flushleft}
\hspace{50bp}$e^{\pm f(x)}=I_{0}(f(x))+2\stackrel[n=1]{\infty}{\sum}(\pm1)^{n}I_{n}(f(x))$,\hspace{85bp}(14
.a)
\par\end{flushleft}

\begin{flushleft}
\hspace{50bp}$e^{\pm f(x)cos(2\theta)}=I_{0}(f(x))+2\stackrel[n=1]{\infty}{\sum}(\pm1)^{n}I_{n}(f(x))cos(2n\theta)$,\hspace{22bp}(14
.b)
\par\end{flushleft}

where $I_{n}(z)$ are the modified Bessel functions of the first kind,
we can rewrite the Eq. (13) as

\begin{flushleft}
\hspace{50bp}$\frac{1}{\sigma}\{\sigma^{\prime\prime}+\sigma^{\prime}(\frac{2}{r}-\mu^{\prime})\}-\lambda e^{\nu+\mu}\sigma^{2}\phi^{2}=\frac{e^{2\mu}}{\phi}(\ddot{\phi}+\overset{.}{\mu}\dot{\phi})$,\hspace{70bp}(15)
\par\end{flushleft}

This equation is not separable due to the second term in left-hand
side. Right-hand side term suggests that the scalar field oscillates
harmonically in time with a damping term related to $\overset{.}{\mu}$.
Following the work {[}8-9{]}, we just consider that

\begin{flushleft}
\hspace{110bp}$\sqrt{k_{0}}\Phi(t,r)=2\sigma(r)cos(\omega t)$,\hspace{96bp}(16)
\par\end{flushleft}

where $\omega$ is the fundamental frequency of the scalar oscillaton.
Integrating on Eq. (7) is a straight forward for obtaining the following
one

\begin{flushleft}
\hspace{100bp}$\nu+\mu=(\nu+\mu)_{0}+r\sigma\sigma^{\prime}cos(2\omega t)$,\hspace{82bp}(17)
\par\end{flushleft}

with $(\nu+\mu)_{0}$ as an arbitrary function of $r-$coordinate
only. Then the metric functions can be expanded as

\begin{flushleft}
\hspace{100bp}$\nu(t,r)=\nu_{0}(r)+\nu_{1}(r)cos(2\omega t)$,\hspace{79bp}(18
.a)
\par\end{flushleft}

\begin{flushleft}
\hspace{100bp}$\mu(t,r)=\mu_{0}(r)+\mu_{1}(r)cos(2\omega t)$,\hspace{76bp}(18
.b)
\par\end{flushleft}

comparison of these two recent equations with Eq. (17) reveals that

\begin{flushleft}
\hspace{130bp}$\nu_{1}+\mu_{1}=r\sigma\sigma^{\prime}$.\hspace{117bp}(18
.c)
\par\end{flushleft}

Then the metric functions can be expanded by using Eq. (14.b) as

\begin{flushleft}
\hspace{30bp}$e^{\nu+\mu}=e^{\nu_{0}+\mu_{0}}[I_{0}(\nu_{1}+\mu_{1})+2\stackrel[n=1]{\infty}{\sum}I_{n}(\nu_{1}+\mu_{1})cos(2n\omega t)]$
\par\end{flushleft}

\begin{flushleft}
\hspace{53bp}$=e^{\nu_{0}+\mu_{0}}[I_{0}(r\sigma\sigma^{\prime})+2\stackrel[n=1]{\infty}{\sum}I_{n}(r\sigma\sigma^{\prime})cos(2n\omega t)]$,\hspace{60bp}(19
.a)
\par\end{flushleft}

\begin{flushleft}
\hspace{33bp}$e^{\nu-\mu}=e^{\nu_{0}-\mu_{0}}[I_{0}(\nu_{1}-\mu_{1})+2\stackrel[n=1]{\infty}{\sum}I_{n}(\nu_{1}-\mu_{1})cos(2n\omega t)]$.\hspace{33bp}(19
.b)
\par\end{flushleft}

These equations show that metric coefficients oscillate in time with
even-multiples of $\omega$ , while scalar field oscillates with odd-multiples
of $\omega$. \vspace{1cm}

\textbf{A. Differential equations\bigskip{}
}

Similar to what has been done for boson star cases in works {[}6,
7, 11{]}, we perform variable changes for numerical purposes of the
following form

\begin{flushleft}
\hspace{45bp}$x=m_{\Phi}r$ , \hspace{15bp}$\varOmega=\frac{\omega}{m_{\Phi}}$
,\hspace{15bp}$e^{\nu_{0}}\rightarrow$$\varOmega e^{\nu_{0}}$ ,\hspace{15bp}
$e^{\mu_{0}}\rightarrow\varOmega^{-1}e^{\mu_{0}}$,\hspace{24bp}(20)
\par\end{flushleft}

where now the metric coefficients are given by $g_{tt}=-\varOmega^{-2}e^{\nu-\mu}$
and $g_{rr}=e^{\nu+\mu}$. It is seen that the mass of scalar field
($m_{\Phi}$) plays a basic role in rescaling of time and distance.
Hence the differential equations for metric functions are obtained
easily from Eqs. (8-11) if we use Eqs. (19- 20) , the scalar field
(16), and setting each Fourier component to zero.

\begin{flushleft}
$\nu_{0}^{\prime}=x[e^{2\mu_{0}}\sigma^{2}\left(I_{0}(2\mu_{1})-I_{1}(2\mu_{1})\right)+\sigma^{\prime2}]$,\hspace{149bp}(21)
\par\end{flushleft}

\begin{flushleft}
$\nu_{1}^{\prime}=x[e^{2\mu_{0}}\sigma^{2}\left(2I_{1}(2\mu_{1})-I_{0}(2\mu_{1})-I_{2}(2\mu_{1})\right)+\sigma^{\prime2}]$,\hspace{100bp}(22)
\par\end{flushleft}

\begin{flushleft}
$\mu_{0}^{\prime}=\frac{1}{x}\{1+e^{\nu_{0}+\mu_{0}}[\frac{1}{2}x^{2}\sigma^{4}\left(3I_{0}(x\sigma\sigma^{\prime})+2I_{1}(x\sigma\sigma^{\prime})+I_{2}(x\sigma\sigma^{\prime}\right)-I_{0}(x\sigma\sigma^{\prime})]\}$,
(23)
\par\end{flushleft}

\begin{flushleft}
$\mu_{1}^{\prime}=\frac{1}{x}e^{\nu_{0}+\mu_{0}}[x^{2}\sigma^{4}\left(2I_{0}(x\sigma\sigma^{\prime})+3.5I_{1}(x\sigma\sigma^{\prime})+2I_{2}(x\sigma\sigma^{\prime})\right)-2I_{1}(x\sigma\sigma^{\prime})]$,\hspace{11bp}(24)
\par\end{flushleft}

\begin{flushleft}
$\sigma^{\prime\prime}=-\sigma^{\prime}(\frac{2}{x}-\mu_{0}^{\prime}-\frac{1}{2}\mu_{1}^{\prime})+\sigma^{3}e^{\nu_{0}+\mu_{0}}[3I_{0}(x\sigma\sigma^{\prime})+4I_{1}(x\sigma\sigma^{\prime})+I(x\sigma\sigma^{\prime})]-e^{2\mu_{0}}\sigma[I_{0}(2\mu_{1})(1-\mu_{1})+I_{1}(2\mu_{1})+\mu_{1}I_{12}(2\mu_{1})]$,\hspace{123bp}(25)
\par\end{flushleft}

where now the primes denote $\frac{d}{dx}$. Meanwhile these equations
are obtained due to rescaling mentioned in Eq. (20) which causes the
following changes in the metric functions $\nu$, $\mu$ and the radial
part of scalar field $\sigma$.

\begin{flushleft}
$\nu(t,r)\equiv\nu(t,x)\rightarrow\nu^{\prime}(t,r)=m_{\Phi}\nu^{\prime}(t,x)$,\hspace{142bp}(26
.a)
\par\end{flushleft}

\begin{flushleft}
$\mu(t,r)\equiv\mu(t,x)\rightarrow\mu^{\prime}(t,r)=m_{\Phi}\mu^{\prime}(t,x)$,\hspace{139bp}(26
.b)
\par\end{flushleft}

\begin{flushleft}
$\sigma(r)\equiv\sigma(x)$$\rightarrow$$\sigma^{\prime}(r)=m_{\Phi}\sigma^{\prime}(x)$$\rightarrow$$\sigma^{\prime\prime}(r)=m_{\Phi}^{2}\sigma^{\prime\prime}(x)$,\hspace{91bp}(26
.c)
\par\end{flushleft}

then constraints imposed to the Eqs.21-25 require we put the condition

\begin{flushleft}
\hspace{130bp}$\lambda=m_{\Phi}^{2}k_{0}$.\hspace{137bp}(26 .d)
\par\end{flushleft}

It is necessary to state that in making the expansions (21-25) the
neglected terms on the right hand side were those containing $cos(4\omega t)$,
$cos(6\omega t)$ and so on, while the neglected ones in Klein-Gordon
equation were those with $cos(3\omega t)$ , $cos(5\omega t)$, and
so on. This suggests that the metric coefficients should be expanded
with even Fourier terms and the scalar field expansion involves only
odd Fourier terms. Then, the expansions used in {[}12{]} are well
justified. By solving equations (21-25) numerically, the solutions
are completely determined then metric functions and metric coefficients
are obtained as well as oscillaton mass and frequency. Before doing
any calculation on these equations, it is recalled that Eq. (18.c)
is \textit{an exact algebraic relation. }This means that we can solve
a system of four ordinary differential equations instead of a system
with five equations\textit{.}

\vspace{1cm}

\textbf{B. Initial Conditions }

\textbf{\bigskip{}
}

Non-singular solutions for a scalar filed at $x=0$ require that $\sigma^{\prime}(0)=0$
and $\nu(t,0)+\mu(t,0)=0,$ so $\nu_{0}(0)=-\mu_{0}(0)$ and $\mu_{1}(0)=-\nu_{1}(0)$.
The latter condition is obtained in shorter way, Eq. (18.c). If the
scalar field vanishes when $x\rightarrow\infty$ then Eq. (16) implies
that $\sigma(\infty)=0$. Asymptotically flatness, complying with
the Minkowski condition, at infinity, requires that $\mu_{1}(\infty)=0$
as $\nu_{1}(\infty)=0,$ but $\mu_{0}(\infty)=-\nu_{0}(\infty)\neq0$
because of the change of variables in (20), and $exp(\nu-\mu)(\infty)=\varOmega^{-2}$
gives the value of $\omega$ (fundamental frequency), while still
$exp(\nu+\mu)(\infty)=1$ {[}6, 7, 11{]}. Now the first step is to
choose a value for $\sigma(0)$ which is called the \textit{central}
value because for each value of $\sigma(0)$ we only have two degrees
of freedom and we need to adjust the central values $\mu_{0}(0)$,
$\mu_{1}(0)$ and then $\nu_{0}(0)$, $\nu_{1}(0)$ results from $\mu$$_{0}(0)$,
$\mu_{1}(0)$. These values are sufficient to obtain different n-nodes
solutions. On the other hand as we can see from Eq. (24), It is important
to mention that the radial derivative of $\mu_{1}(x)$ is always positive.
Hence providing $\mu_{1}(0)<0$ asymptotically flat condition is reached.
On the other hand we have neglected higher terms of expansion (19
-a) and (19 -b), therefore the condition $|\mu_{1}|<1$ is needed
for the solutions of Eqs. (21-25) to converge. \vspace{1cm}

\textbf{C. Numerical results\bigskip{}
}

If we expand the metric as

\begin{flushleft}
\hspace{8bp}$g=g_{0}(x)+g_{2}(x)cos(2\omega t)+g_{4}(x)cos(4\omega t)+...=\stackrel[n=0]{\infty}{\sum}g_{2n}(x)cos(2n\omega t)$,\hspace{12bp}(27)
\par\end{flushleft}

and comparing it with Eqs. (19 a, 19 b) then, typical metric coefficients
for 0-node solution are obtained. The radial and time metric coefficients
with a central value of $\sigma(x=0)=0.4$ and other boundary conditions
are shown in Fig. 1.

\begin{figure}
\includegraphics[width=10cm,height=8cm]{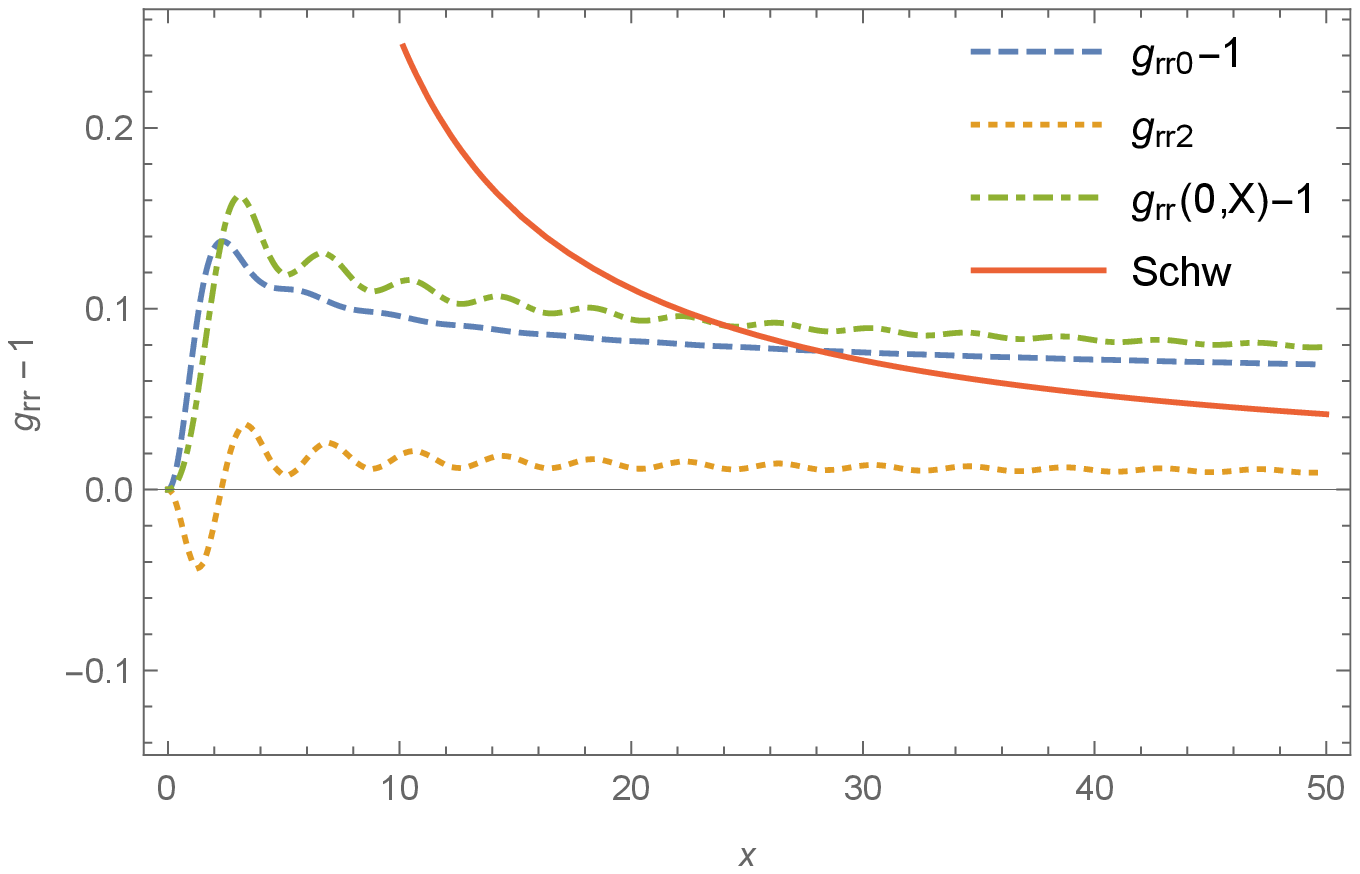}

\includegraphics[width=10cm,height=8cm]{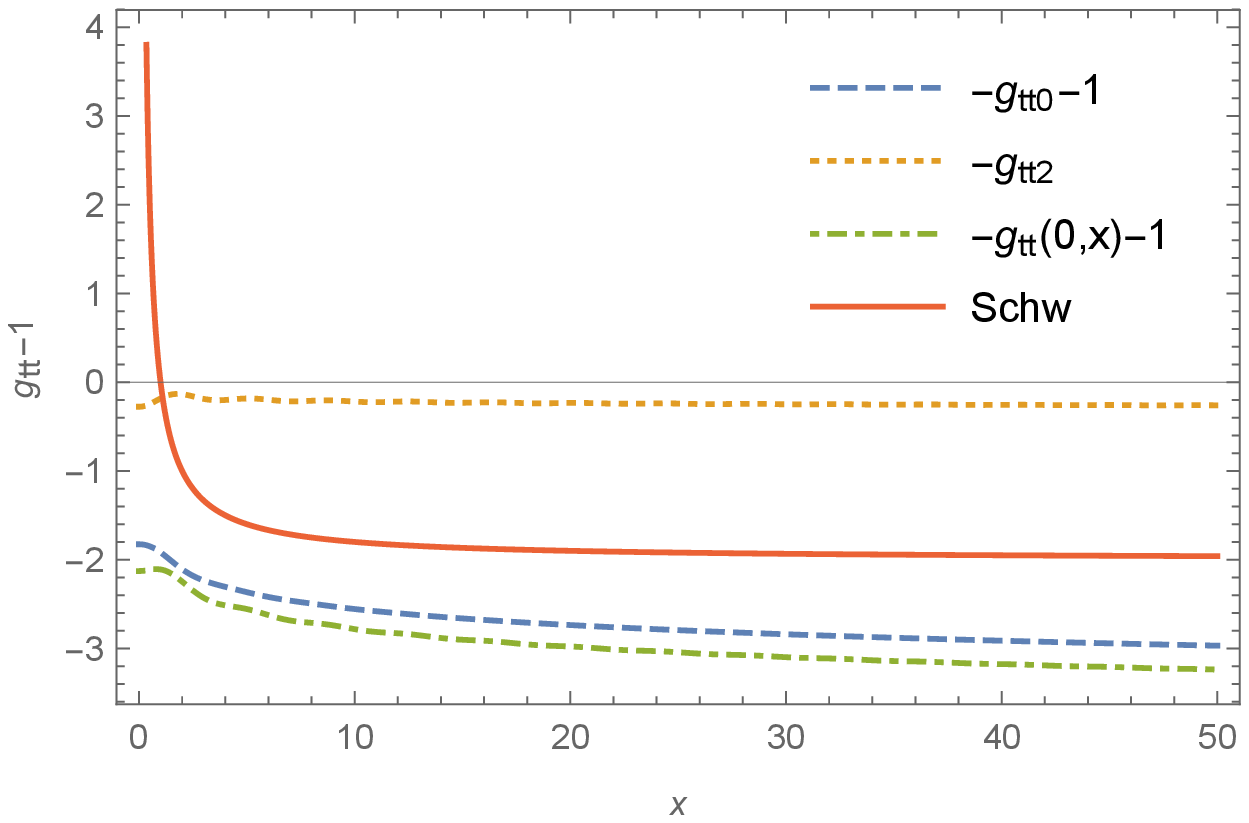}

\caption{Metric coefficients $g_{rr}-1$ and $-g_{tt}-1$ and their respective
two first terms, $g_{rr0}-1$, $g_{rr2}$ (top) and $-g_{tt0}-1$,
$-g_{tt2}$ (bottom) for a central value $\sigma(x=0)=0.4,$ $\nu_{0}(x=0)=-0.11$,
$\nu_{1}(x=0)=0.17,$ $\mu_{0}(x=0)=0.11,$ $\mu_{1}(x=0)=-0.17$
according to explained in the text and the Schwarzschild metric coefficients
are also shown for the solution of the same mass.}
\end{figure}

In Figs. 2. and 3. the answers for differential equations mentioned
by Eqs. (21-25) are shown.

As we can see from Fig. 3. the radial part of scalar field, $\sigma$,
with a damping decrease becomes negative for some values of $x$ .
This means that negative scalar fields have an effective role in being
of oscillatons.

\medskip{}

\begin{figure}
\includegraphics[width=10cm,height=8cm]{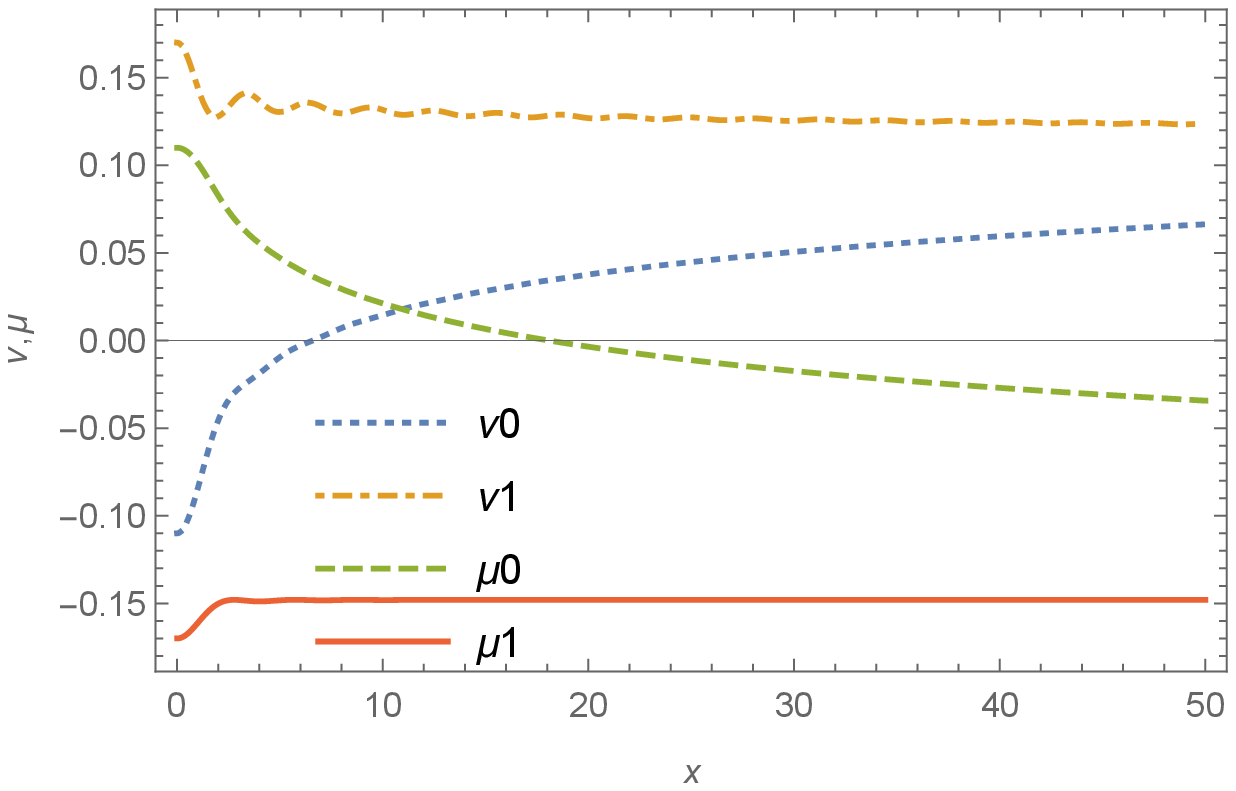}\caption{Metric functions $\nu_{0},$ $\nu_{1},$ $\mu_{0}$ and $\mu_{1}$with
the initial boundary conditions mentioned in Fig. 1.}
\end{figure}

\begin{figure}
\includegraphics[width=10cm,height=8cm]{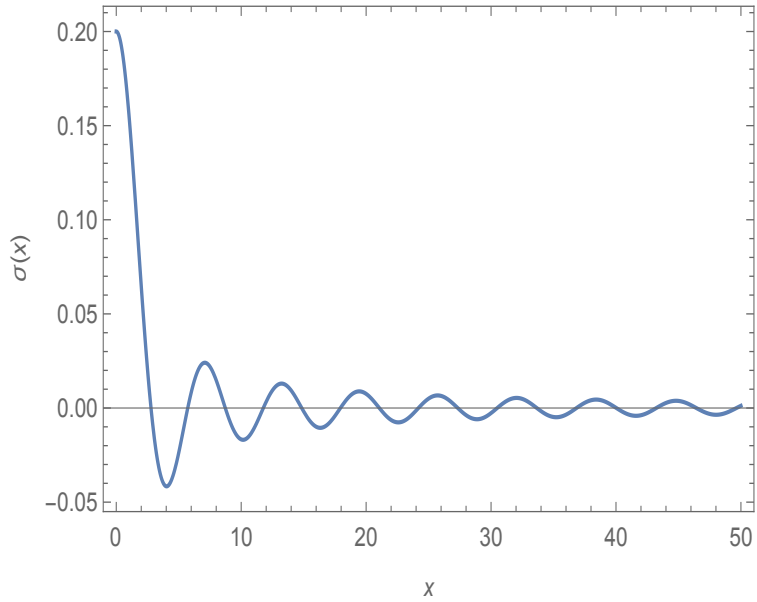}\caption{Radial part of the scalar field with the boundary conditions mentioned
in Fig. 1 and 2.}
\end{figure}

Similar to what has been done for metric coefficient, we can rewrite
the energy density, Eq. (3), the radial pressure, Eq. (5) and the
angular pressure, Eq. (6) for oscillaton as

\begin{flushleft}
$\rho_{\Phi}(t,x)=\frac{1}{8\pi}m_{pl}^{2}m_{\Phi}^{2}\{\sigma^{2}e^{-(\nu-\mu)}[1-cos(2\omega t)]+\sigma^{\prime2}e^{-(\nu+\mu)}[1+cos(2\omega t)]+\frac{\sigma^{4}}{2}[cos(4\omega t)+4cos(2\omega t)+3]\}$
\hspace{193bp}(28)
\par\end{flushleft}

\begin{flushleft}
$p_{r}(t,x)=\frac{1}{8\pi}m_{pl}^{2}m_{\Phi}^{2}\{\sigma^{2}e^{-(\nu-\mu)}[1-cos(2\omega t)]+\sigma^{\prime2}e^{-(\nu+\mu)}[1+cos(2\omega t)]-\frac{\sigma^{4}}{2}[cos(4\omega t)+4cos(2\omega t)+3]\}$
\hspace{193bp}(29)
\par\end{flushleft}

\begin{flushleft}
$p_{\bot}(t,x)=\frac{1}{8\pi}m_{pl}^{2}m_{\Phi}^{2}\{\sigma^{2}e^{-(\nu-\mu)}[1-cos(2\omega t)]-\sigma^{\prime2}e^{-(\nu+\mu)}[1+cos(2\omega t)]-\frac{\sigma^{4}}{2}[cos(4\omega t)+4cos(2\omega t)+3]\}$
\hspace{193bp}(30)
\par\end{flushleft}

Equations 28-30 show that different nodes of energy density, radial
and angular components of pressure can be obtained easily through
their expansion. The values of $\rho_{\Phi}$ for times $\omega t=0$,
$\frac{\pi}{2}$ and $\rho_{\Phi0}$ using the Fourier expansion to
second order are shown in Fig. 4.

\begin{figure}
\includegraphics[width=10cm,height=8cm]{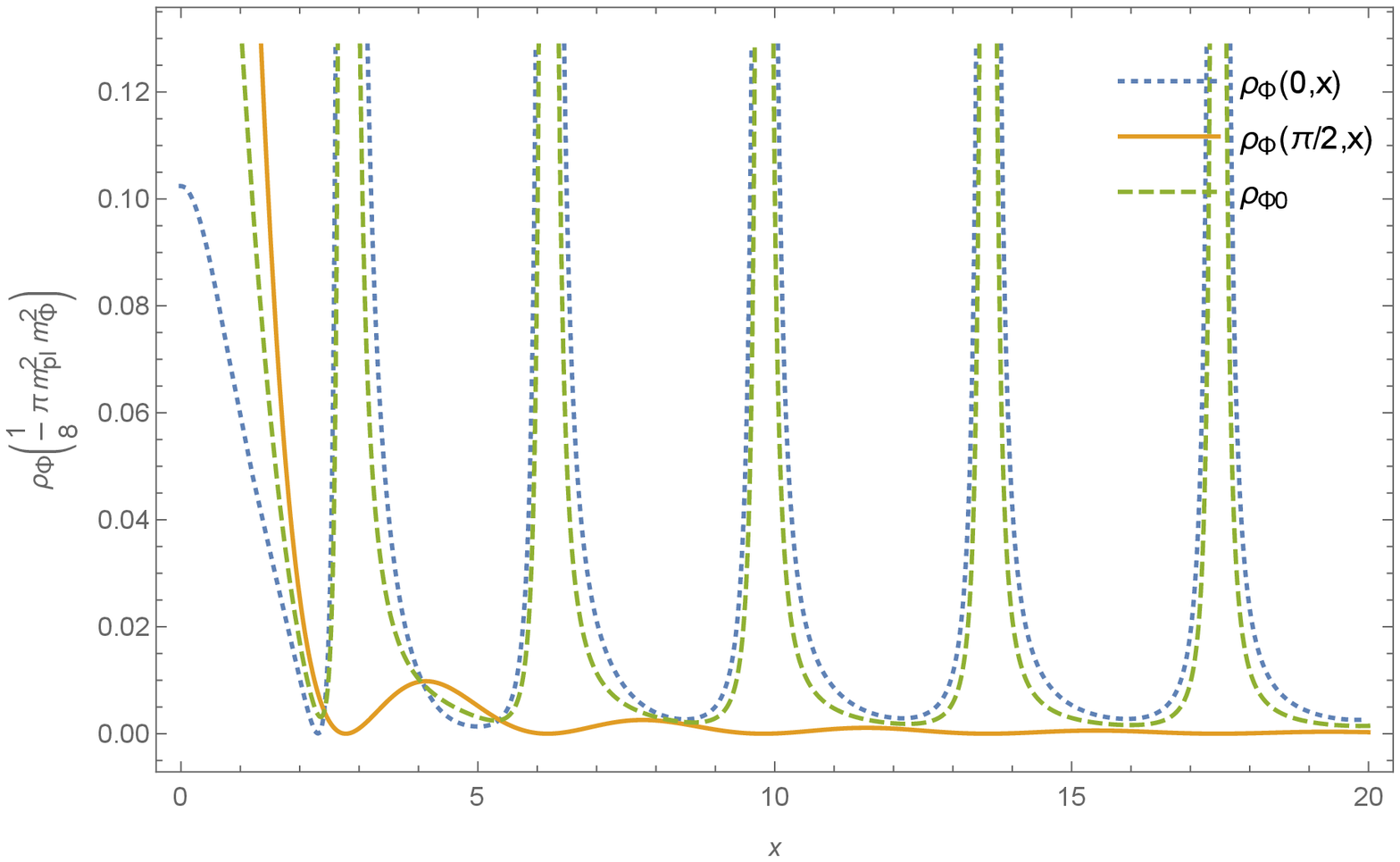}\caption{The energy density function $\rho_{\Phi}(\omega t,x)$ (see Eq. 28)
for the oscillaton in Fig. 1, 2.}
\end{figure}

The values calculated for the components of radial and angular pressure
for times $\omega t=0,\frac{\pi}{2}$ and zero nodes are shown in
Fig. 5. As we can see from Fig. 5. both radial and angular components
have a negative effective pressure for some values of $x$. This means
that for negative pressure work is done on the oscillaton when it
expands. On the other hand by using Eq . (4) and (16) we can also
evaluate the momentum density of the oscillaton. The values of $p_{\Phi}$
for times $\omega t=0,\frac{\pi}{2}$ are shown in Fig. 5. As we can
see from Figs.4 and 5.
\begin{figure}
\includegraphics[width=10cm,height=8cm]{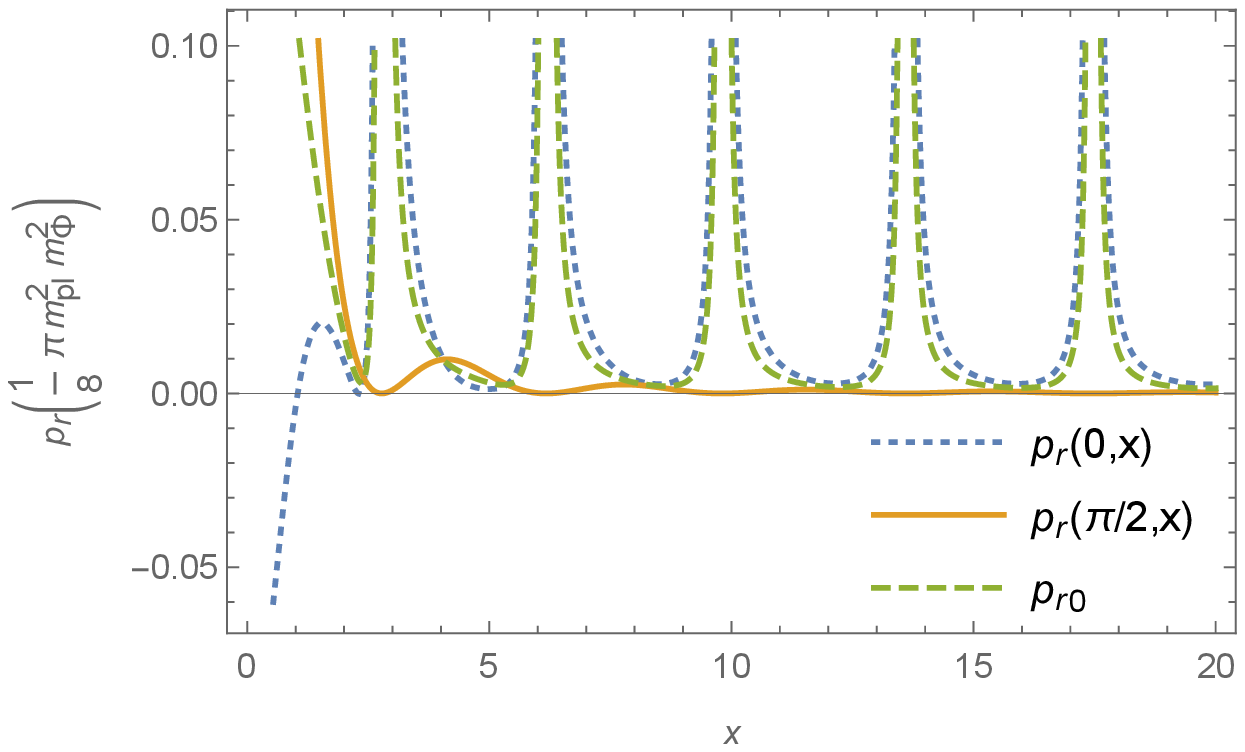}

\includegraphics[width=10cm,height=8cm]{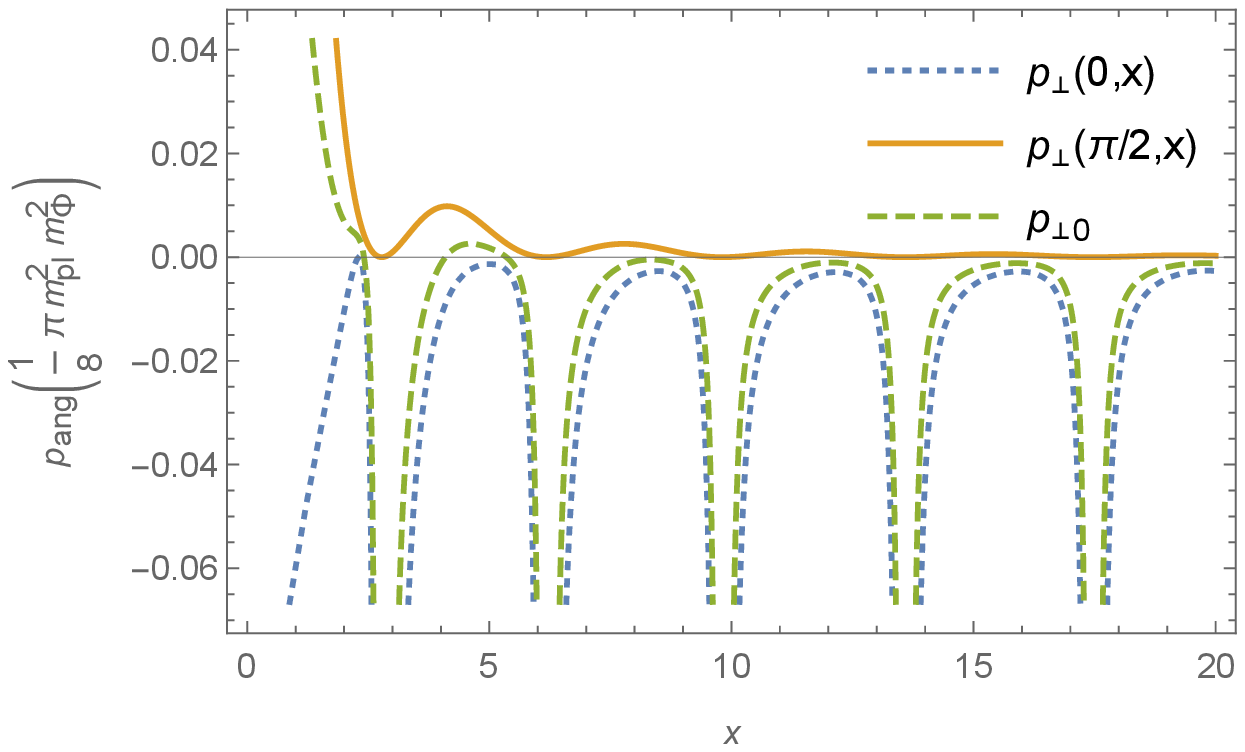}\caption{The radial pressure function $p_{r}(\omega t,x)$ (see Eq. 29) (top)
and angular pressure function $p_{\bot}(\omega t,x)$ (see Eq. 30
) (bottom) for the solutions in Fig. 1, 2.}
\end{figure}

\begin{figure}
\includegraphics[width=10cm,height=8cm]{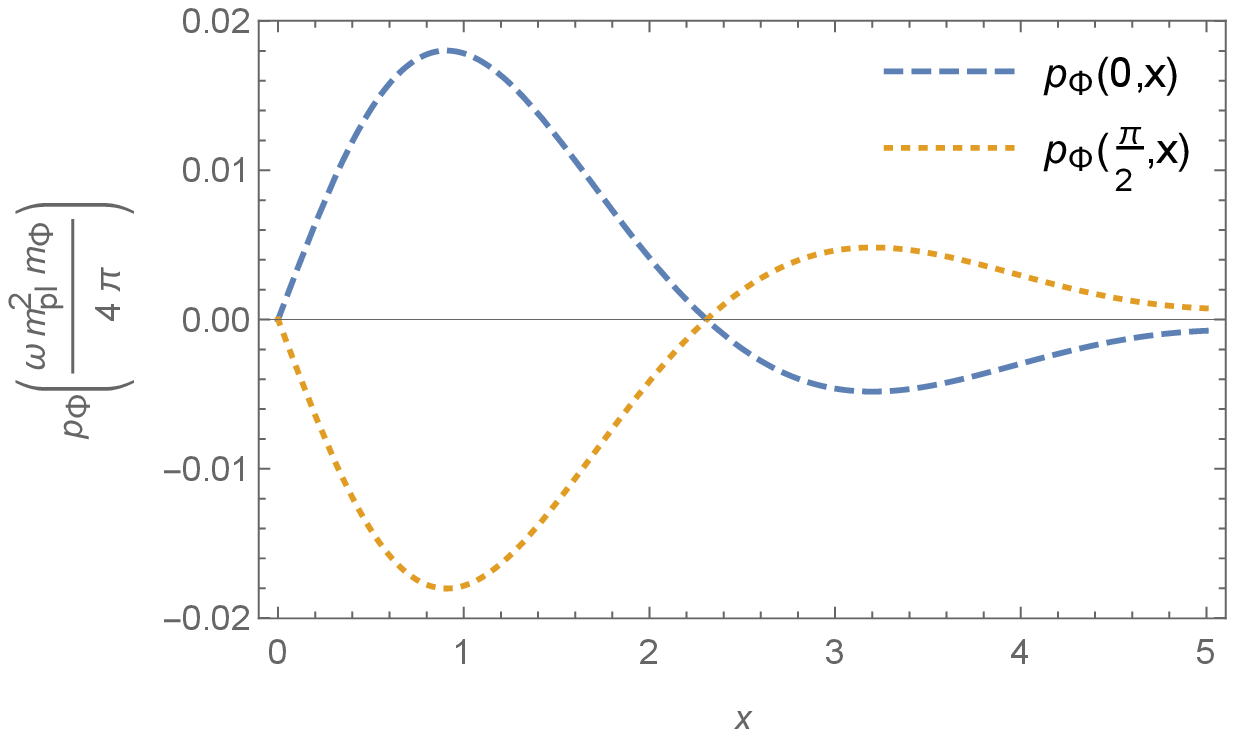}

\caption{The momentum density function for the solutions in Fig. 1, 2.}
\end{figure}

Since the metric coefficients are asymptotically flat, static and
comply with their corresponding ones in the Minkowski situation when
$x\rightarrow\infty$, therefore $exp(\nu+\mu)$ is identified as
$\frac{1}{1-\frac{2GM_{\Phi}}{r}}$ . Then the mass seen by an observer
at infinity may be calculated as {[}7-9{]}

\begin{flushleft}
\hspace{100bp}$M_{\Phi}=(\frac{m_{pl}^{2}}{m_{\Phi}})\underset{x\rightarrow\infty}{lim}\frac{x}{2}(1-e^{-\nu-\mu})$.\hspace{90bp}(31)
\par\end{flushleft}

This is the mass which is related to the scalar field and can be employed
as a possibility for the role of dark matter. Equation (31) shows
that the calculated mass of the oscillaton is constant which means
that the masses observed at infinity are the same for all times. This
is natural because oscillaton should be in line with the Schwarzschild
solution for the same mass according to Birkhoff$^{,}$s theorem{[}13{]}.
Here some thing is unusual, and that is, for $\sigma(x=0)<0.235$
the mass obtained from the Eq. (31) is negative. We can see that there
is a negative maximum mass $M_{max}=-0.23377m_{Pl}^{2}/m$ with $\sigma_{c}(x=0)=0.175.$
Whereas the $p_{r}(0,x)$ (main component of radial pressure) at least
for these mentioned values, $\sigma(x=0)<0.235,$ are negative. The
closest known real representative of such exotic matter is a region
of pseudo-negative pressure density produced by the \textit{Casimir}
effect {[}14-17{]}. Therefore the\textit{ negative mass} can be described
by this model, quartic potential, the advantage which distinguishes
this model from quadratic and exponential scalar potential. For values
higher than $\sigma(x=0)>0.235$ the mass values are positive and
increase rapidly.

\begin{figure}
\includegraphics[width=10cm,height=5cm]{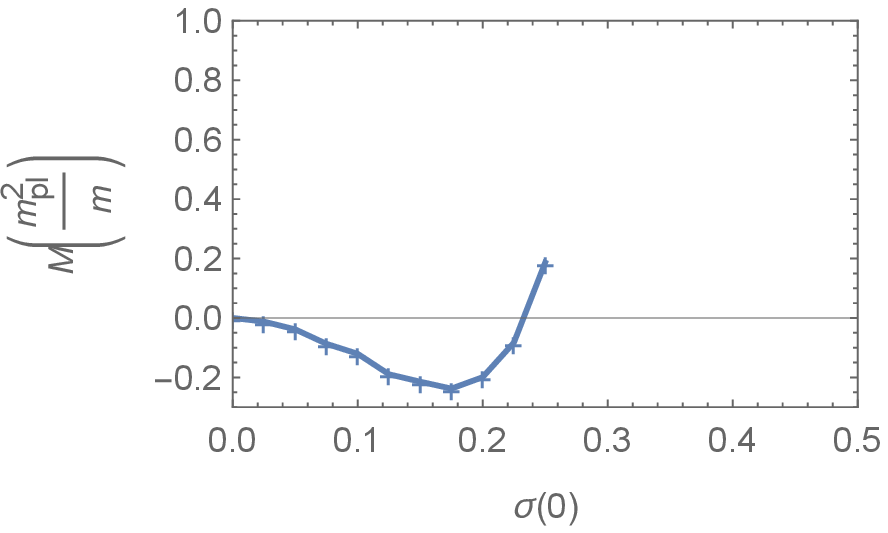}

\includegraphics[width=10cm,height=8cm]{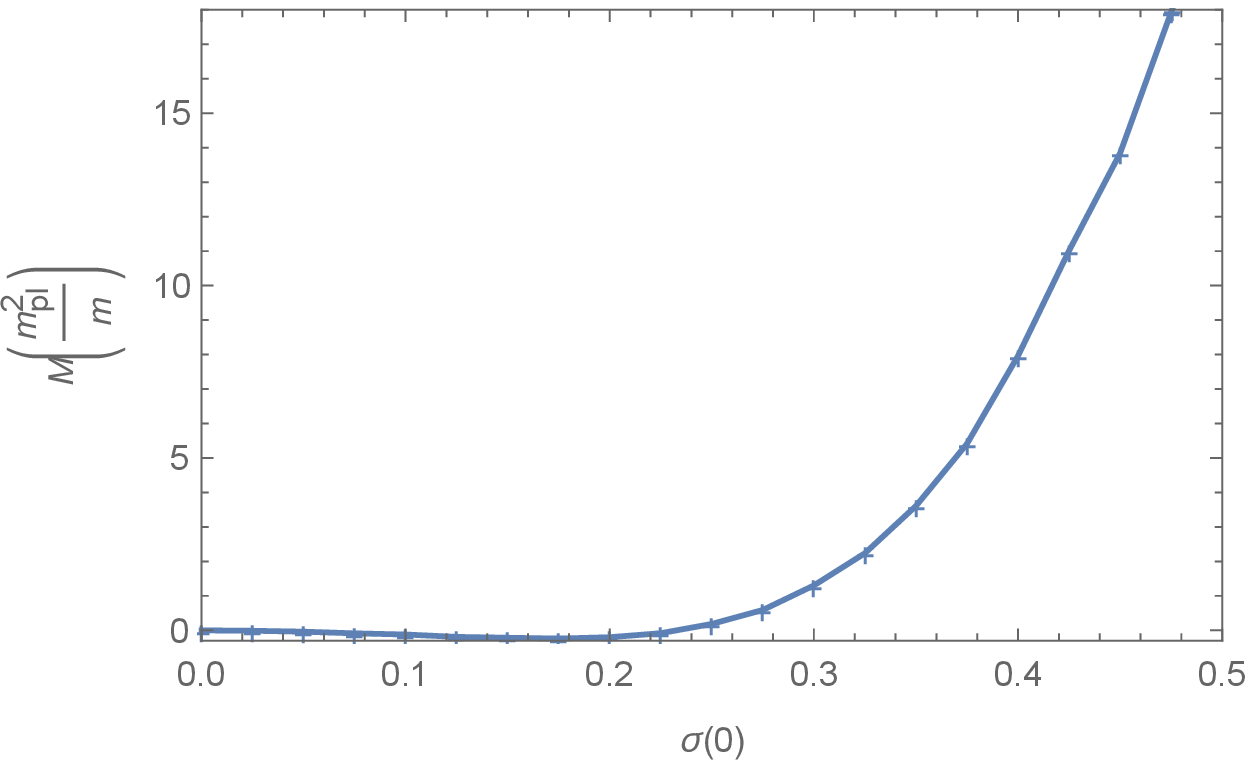}

\caption{The mass observed at infinity (equation 31) for different central
values of $\sigma(0)$.}
\end{figure}

As we mentioned in boundary conditions, the fundamental frequency
is obtained by asymptotic value $exp(\nu-\mu)(\infty)=\Omega^{-2}$.
In that we had $\mu_{0}(\infty)=-\nu_{0}(\infty)\neq0$ due to the
boundary conditions and taking into account the rapid convergence
of $\nu_{0}$ (Fig. 2), then we can have

\hspace{128bp}$\Omega=e^{-\nu_{0}(\infty)}$.\hspace{126bp}(32)

The profile of the fundamental frequencies are shown in Fig. 8. It
is clear that more massive oscillatons oscillate with smaller frequencies.\bigskip{}

\begin{figure}
\includegraphics[width=10cm,height=8cm]{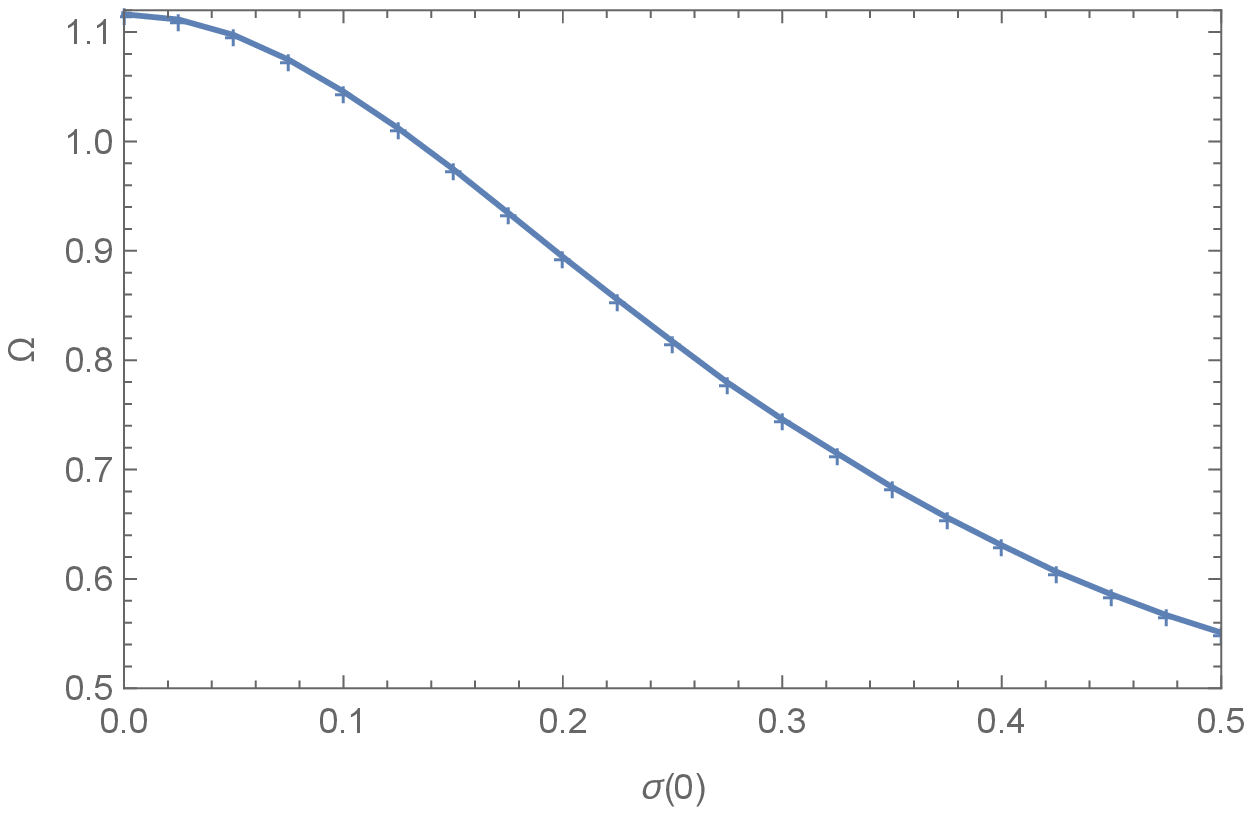}

\caption{The fundamental frequencies $\Omega$ for corresponding masses shown
in Fig. 7.}
\end{figure}

\textbf{IV. THE STATIONARY LIMIT PROCEDURE}

For weak field condition in which $\sigma(0)\ll1$ (in consequence
|$\mu_{1}|\ll1$ ), It is easy to simplify the Eqs. (21-25) as far
as possible as

\bigskip{}

\hspace{110bp}$\nu_{0}^{\prime}=x[e^{2\mu_{0}}\sigma^{2}+\sigma^{\prime2}]$,\hspace{110bp}(33)

\bigskip{}

\hspace{110bp}$\mu_{0}^{\prime}=\frac{1}{x}\{1+e^{\nu_{0}+\mu_{0}}(\frac{3}{2}x^{2}\sigma^{4}-1)\},$\hspace{63bp}(34)

\bigskip{}

\hspace{110bp}$\mu_{1}^{\prime}=\sigma e^{\nu_{0}+\mu_{0}}(x^{2}-1)$,\hspace{106bp}(35)

\bigskip{}

\hspace{110bp}$\sigma^{\prime\prime}=-\sigma^{\prime}(\frac{2}{x}-\mu_{0}^{\prime})-3\sigma^{3}e^{\nu_{0}+\mu_{0}}-\sigma e^{2\mu_{0}},$\hspace{34bp}(36)

\vspace{0.5cm}

where for $z\ll1$, we have used this fact that $I_{0}(z)\sim\mathcal{O}(1)$,
$I_{1}(z)\sim\mathcal{O}(\frac{z}{2})$ and higher orders, $I_{n}(z)$
, are neglected, while Eq. (18-c) with regard to variable changes
mentioned in Eq. (20) remains without change.

\begin{figure}
\includegraphics[width=10cm,height=8cm]{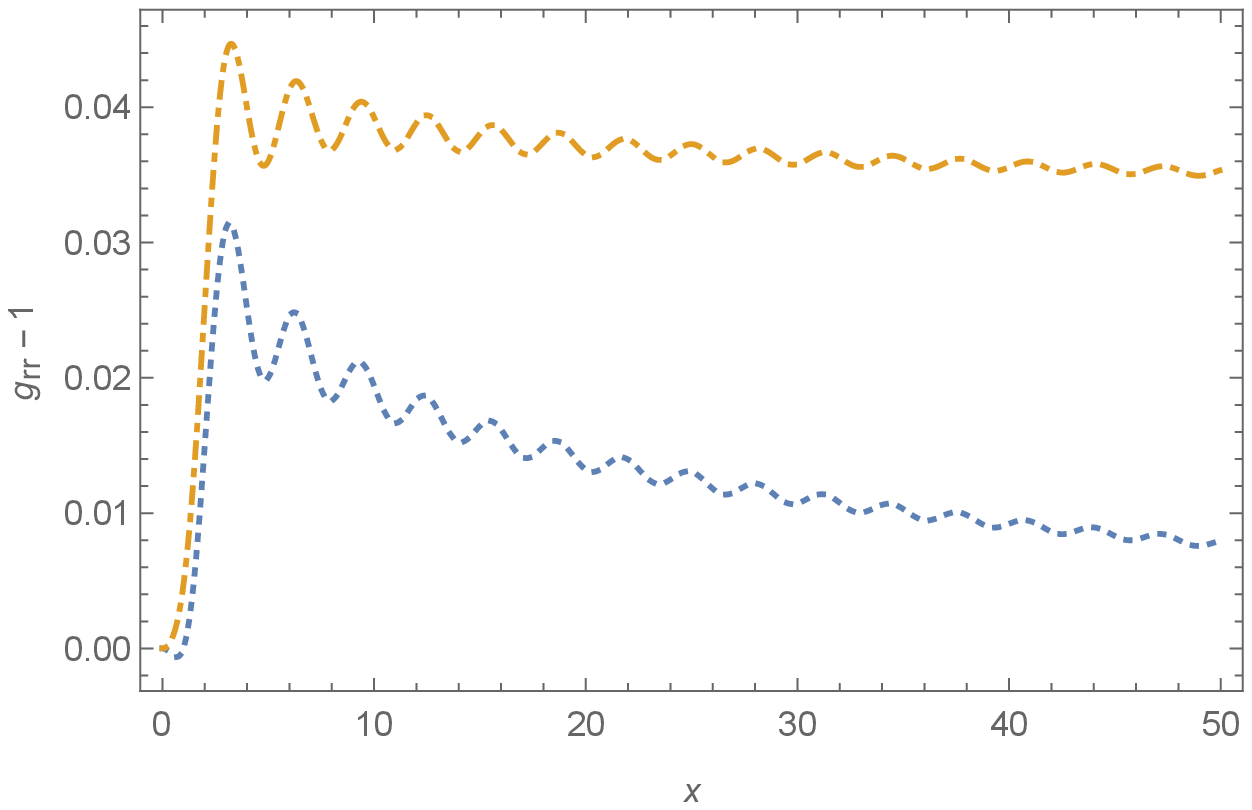}

\includegraphics[width=10cm,height=8cm]{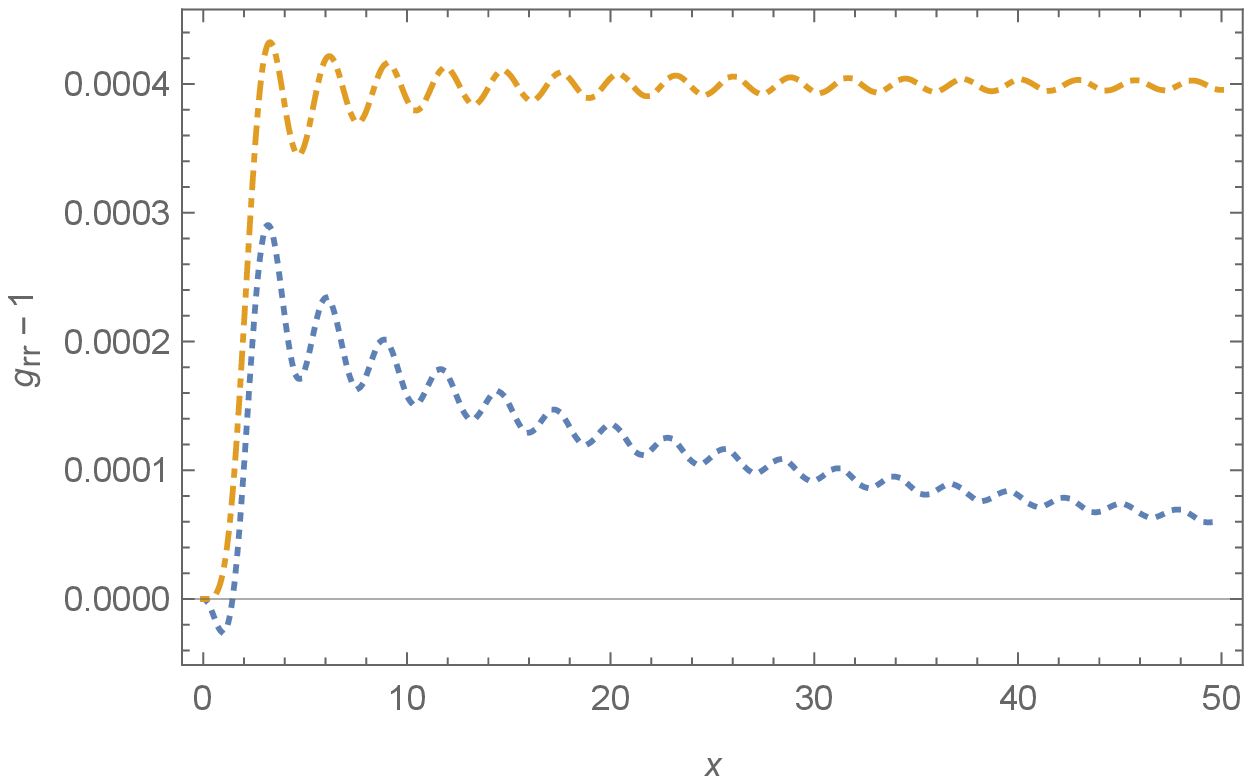}

\caption{Comparison of the numerical results for the metric coefficient $g_{rr}-1$
for a quartic scalar potential oscillaton calculated by the method
of sec.III (dotted) and with the stationary limit procedure (dotDashed)
for two different central values of $\sigma(0)=0.2$ (top) and $\sigma(0)=0.02$
(bottom).}
\end{figure}

If we expand the scalar potential in a Fourier series by taking Eq.
(16)

\begin{flushleft}
$V(\Phi)=\stackrel[n=0]{4}{\sum}V_{n}(\sigma)cos(n\omega t)=\frac{m_{\varPhi}^{2}m_{Pl}^{2}}{16\pi}\sigma^{4}[cos(4\omega t)+4cos(2\omega t)+3]$,\hspace{36bp}(37)
\par\end{flushleft}

then, as we have $\lambda=m_{\Phi}^{2}k_{0}$ , then we can obtain

\hspace{110bp}$\sigma=(\frac{16\pi V_{0}(\sigma)}{m_{Pl}^{2}m_{\varPhi}^{2}})^{\frac{1}{4}}$.\hspace{125bp}(38)

It is clear that Eq. (36) is the only equation which will change among
(33-36). Therefore it can be rewritten as

\hspace{45bp}$\sigma^{\prime\prime}=-\sigma^{\prime}(\frac{2}{x}-\mu_{0}^{\prime})-3(\frac{16\pi V_{0}(\sigma)}{m_{Pl}^{2}m_{\varPhi}^{2}})^{\frac{1}{4}}e^{\nu_{0}+\mu_{0}}-(\frac{16\pi V_{0}(\sigma)}{m_{Pl}^{2}m_{\varPhi}^{2}})e^{2\mu_{0}}.$\hspace{30bp}(39)

At this stage it is necessary to recall that, since oscillatons are
made of real scalar fields, therefore we know from non-relativistic
field theory that charge and current densities which are identified
as $\rho$ and $\overrightarrow{J}$ respectively should be equal
to zero, then these objects are electrically neutral. On the other
hand real $\Phi$ corresponds to electrically neutral particles in
the oscillaton environment, hence we do not expect any electromagnetic
wave emitting from oscillatons {[}9{]}.

Another interesting thing is that: Could the oscillatons predicted
by the scalar field be somehow associated with the ''gravitational
waves'' phenomenon?

As a motivation for the this issue, we start with the following reasoning.
In the actual status of our understanding of the universe, there is
an apparent asymmetry in the kind of interactions that take part in
nature. The Scalar Field Dark Matter Model: A Braneworld Connection
known fundamental interactions are either spin-1, or spin-2. Electromagnetic,
weak and strong interactions are spin-1 interactions, while gravitational
interactions are spin-2. Of course, this could be just a coincidence.
Nevertheless, we know that the simplest particles are the spin-0 ones.
The asymmetry lies in the fact that there is no spin-0 fundamental
interactions. Why did Nature forget to use spin-0 fundamental interactions?
On the other hand, we know from the success of the $\Lambda CDM$
model that two fields currently take the main role in the Cosmos,
the dark matter and the dark energy. Recently, it has been indeed
proposed that dark matter is a scalar field, that is, a spin-0 fundamental
interaction. This is the so called Scalar Field Dark Matter $(SFDM)$
hypothesis . If true, this hypothesis could solve the problem of the
apparent asymmetry in our picture of nature {[}18{]}. As a final part
of this work, it is interesting to do a comparison between these kinds
of oscillatons made by quartic scalar potential and our previous work
which described by exponential scalar potential {[}9{]}. For quartic
scalar potential, metric coefficients, for different values of $\sigma$
, comply with flatness condition asymptotically much more better than
their corresponding ones in exponential scalar potential as well as
frequencies. But in contrast to exponential and quadratic scalar potential
we have several singularity points in energy density, radial and angular
components of pressure in this kind of potential with no persuasive
explanation {[}7, 9{]}.

\vspace{1cm}

\textbf{IV. CONCLUSIONS\bigskip{}
}

In this paper we presented the simplest approximation for solving
the minimally coupled Einstein-Klein-Gordon equations for a spherically
symmetric oscillating soliton object endowed with a scalar quartic
potential field $V(\Phi)=\frac{1}{4}\lambda\varPhi^{4}$ and an harmonic
time-dependent scalar field $\Phi.$ By taking into account the Fourier
expansions of differential equations and with regard to the boundary
conditions which require the non-singularity and asymptotically flatness,
solutions are obtained easily. It should be emphasized that a dynamical
situation is imposed on the region of the oscillaton only, therefore
we have asymptotically static metric and solutions. This fact helps
us to find the mass of these astronomical objects as the most important
topics to justify what called dark matter as well as their fundamental
frequency. Results show that a quartic scalar field potential causes
different profiles for metric functions and metric coefficients as
well as energy density and mass distribution in comparison with what
has been done in previous works for quadratic and exponential scalar
field potentials. On the other hand with the same boundary initial
conditions, all kind of the potentials, have the same fundamental
frequency and mass relations {[}6,7,9{]}. Nevertheless, there are
some more problems that should be investigated for oscillatons derived
from a quartic scalar field. Here are some of these problems:
\begin{itemize}
\item In Fourier expansion, we have used to second order only for simplicity
and higher order requires more complex calculation.
\item For quartic potential studied in this research for $\sigma(x=0)<0.235$
we have negative mass which can be justified by Casimir effect and
negative pressure, but more research should be carried out in this
field.
\item For $\sigma(x=0)>0.325,$ the mass values increase rapidly.
\end{itemize}
\bigskip{}

\end{document}